\documentclass[a4paper,11pt]{article}
\def\today{Submitted 1~December~2011, revised 27~January~2012, accepted 30 January 2012}

\newtheorem{remark}{Remark}

\def \D {\hbox{d}}
\def \Elemsimp {\mathop{\rm H}\nolimits}
\def \mod#1{\vert #1 \vert}

\def \csi{\kappa_{\rm i}} 
\def \ex{e_1} % ( c s_i)^2/48.                     
\def \ey{e_0} %  g_r/36.                           
\def \dlogA{B} % d \log(A e^{i \omega t}) / d \xi. 
\def\Arbzero{M_{0}}
\def\Arbone{M_{1}}
\def\as{M_{1s}}
\def\ad{M_{1d}}
\def\ms{M_{0+}^2 + M_{0-}^2}
\def\md{M_{0+}^2 - M_{0-}^2}
\def \jmax{J}

\def \D {\hbox{d}}

\def \ccomma{\raise 2pt\hbox{,}}

\title{Detection and construction of an elliptic solution
to the complex cubic-quintic Ginzburg-Landau equation\footnote{ 
To appear, \textit{Theoretical and Mathematical Physics}. 
Solitions in 1+1 and 2+1dimensions. DS, KP and all that,
Lecce, 13--14 September 2011.
}
}

\author{\textsc{Robert Conte${}^{1,2}$ and Tuen-Wai Ng${}^{2}$} 
\\ \noindent 1.
LRC MESO, 
Centre de math\'ematiques et de leurs applications (UMR 8536) 
\\ et CEA-DAM, \'Ecole normale sup\'erieure de Cachan, 61, avenue du Pr\'esident Wilson,
\\ F--94235 Cachan Cedex, France.
\smallskip
\\ \noindent 2.
Department of Mathematics,
The University of Hong Kong,
Pokfulam Road.
\smallskip
\\ \noindent E-mail: Robert.Conte@cea.fr, ntw@maths.hku.hk 
}

\date{\today}
\begin{document}
\maketitle


\begin{abstract}
In evolution equations for a complex amplitude,
the phase obeys a much more intricate equation than the amplitude.
Nevertheless, general methods should be applicable to both variables.
On the example of the traveling wave reduction
of the complex cubic-quintic Ginzburg-Landau equation (CGL5),
we explain how to overcome the difficulties
arising in two such methods:
(i) the criterium that the sum of residues of an elliptic solution
should be zero,
(ii) the construction of a first order differential equation admitting the given equation 
as a differential consequence (subequation method).
\end{abstract}

% ==================================
\section{Introduction. Modulus \textit{vs}.~phase in amplitude equations}
\label{sectionIntro}

The time evolution equation $A_t + \cdots=0$ for a complex amplitude $A(x,t)$
is usually, from physical requirements,
invariant under an arbitrary shift of the phase $\varphi=\arg A$,
in which 
$M$ and $\varphi$ denote the modulus and phase, $A=M e^{i \varphi}$.
As a consequence, in the coupled partial differential system for $(M,\varphi)$,
the variable $\varphi$ only contributes by its derivatives.
Then, under a reduction to an ordinary differential equation (ODE)
such as the travelling wave reduction
\begin{eqnarray}
& & 
M \to \tilde{M}(\xi),\ \varphi \to -i \omega t + \tilde{\varphi}(\xi),\ \xi=x-c t,
\end{eqnarray}
in the coupled ODE system for $(\tilde{M},\psi=\tilde{\varphi}')$,
the highest derivation order for $\psi$ will be one less 
than the derivation order for $M$.
Consequently, 
the ODE for $\psi$ obtained by the elimination of $M$
will be much more complicated (by both its volume and its structure of singularities)
than the ODE for $M$ obtained by the elimination of $\psi$.
\medskip

Let us take as an example 
the one-dimensional cubic-quintic complex Ginzburg-Landau equation (CGL5),
\begin{eqnarray}
& &  {\hskip -18.0 truemm}
i A_t +p A_{xx} +q \mod{A}^2 A +r \mod{A}^4 A -i \gamma A =0,\
(A,p,q,r) \in \mathcal{C},\
p r \not=0,\ \Im(r/p)\not=0.
\label{eqCGL5}
\end{eqnarray}
which depends on seven real parameters since $\gamma$ can be chosen real.
For a summary of results on CGL5, see the reviews \cite{AK2002,vS2003}.
Its travelling wave reduction 
\begin{eqnarray}
& & {\hskip -10.0 truemm}
A(x,t)=\sqrt{M(\xi)} e^{i(\displaystyle{-\omega t + \varphi(\xi)})},\
\xi=x-ct,\
(c,\omega,M,\varphi) \in {\mathcal R}.
\label{eqCGL35red}
\\ & &
{\hskip -10.0 truemm}
 \frac{M''}{2 M} -\frac{{M'}^2}{4 M^2} + i \varphi''- {\varphi'}^2
        + i \varphi' \frac{M'}{M}
- i \frac{c}{2p} \frac{M'}{M}
+ \frac{c}{p} \varphi' 
+ \frac{q}{p} M
+ \frac{r}{p} M^2
+ \frac{\omega - i \gamma}{p}
=0,
\label{eqCGL5ReducComplex}
\end{eqnarray}
introduces two additional real constants $(c,\omega)$,
but the total number of real parameters (nine) can be lowered to seven
by a translation of $\varphi'$ and by noticing 
\cite{vS2003}
that the velocity $c$ and the imaginary part of $1/p$
only contribute by their product.
Indeed, denoting the eight real parameters as 
$e_r,e_i,d_r,d_i,s_r,s_i,g_r,g_i$,
\begin{eqnarray}
& &
e_r + i e_i = \frac{r}{p},\
d_r + i d_i = \frac{q}{p},\
s_r - i s_i = \frac{1}{p},\
g_r + i g_i=\frac{\gamma + i \omega}{p}
 + \frac{c^2 s_r}{4} (2 s_i + i s_r),
\end{eqnarray}
and performing the translation
\begin{eqnarray}
& &
\varphi' = \frac{c s_r}{2} + \psi,\
\label{eqdefpsi}
\end{eqnarray}
the system only depends on 
the seven real parameters $e_r,e_i,d_r,d_i,g_r,g_i,c s_i \equiv \csi$.
\medskip

The coupled two-component system in the real variables $(M,\psi)$,
\begin{eqnarray}
& &
\left\lbrace
\begin{array}{ll}
\displaystyle{
\frac{M''}{2 M} -\frac{{M'}^2}{4 M^2} - \csi \frac{M'}{2 M} 
- \psi^2+ e_r M^2+ d_r M + g_i =0,
}\\ \displaystyle{
\psi' + \psi \frac{M'}{M} - \csi \psi + e_i M^2 + d_i M - g_r =0,
}
\end{array}
\right.
\label{eqCGL5ReducRealSystem}
\end{eqnarray}
contains as highest derivatives $M''$ and $\psi'$.
The following parity invariances of (\ref{eqCGL5ReducRealSystem})
\begin{eqnarray}
& q=0:\ & (M,\psi,\xi) \to (-M,\psi,\xi),
\label{eqInvariance_q}
\\
& \csi=0:\ & (M,\psi,\xi) \to (M,-\psi,-\xi),
\label{eqInvariance_csi}
\end{eqnarray}
\medskip
will be used later on.
\medskip

The elimination of $\psi$ yields a one-line 
real third order second degree ODE for $M$
\cite{Klyachkin}
\begin{eqnarray}
& &
\psi = \frac{2 \csi G - G'}{2 M^2 (e_i M^2 + d_i M - g_r)},\ 
\psi^2=\frac{G}{M^2},
\label{eqCGL5Phiprime}
\\
& &
(G'-2 \csi G)^2 - 4 G M^2  (e_i M^2 + d_i M - g_r)^2=0,\ 
\label{eqCGL5Order3}
\\
& &
G=\frac{1}{2} M M'' - \frac{1}{4} M'^2
  -\frac{\csi}{2} M M' + e_r M^4 + d_r M^3 + g_i M^2,
\end{eqnarray}
while the elimination of $M$ yields a 
third order fourth degree ODE for $\psi$ which contains 11053 terms,
whose dominant ones (in the sense of singularities,
as developed in section \ref{sectionPZ}) are,
\begin{eqnarray}
& & {\hskip -15.0truemm}
e_i^6 \psi^4
\left\lbrace
\left\lbrack 
25 e_i^3 \psi^2 \psi'''-10 e_i^2 (3 e_i \psi'+4 e_r \psi^2) \psi \psi_2
 -6 {\psi'}^3 e_i^3-24 e_r e_i^2 {\psi'}^2 \psi^2 
\right. \right. \nonumber \\ & & {\hskip -15.0truemm} \left. \left. \phantom{12345}
 +e_i (80 e_i^2-112 e_r^2) \psi^4 \psi'
 -32 e_r (5 e_i^2+e_r^2) \psi^6 
\right\rbrack^2
\right. \nonumber \\ & & {\hskip -15.0truemm} \left. \phantom{12345}  
 -4 \left\lbrack -3 e_i^2 {\psi'}^2 +12  e_r e_i \psi' \psi^2+4 (5 e_i^2+2 e_r^2) \psi^4\right\rbrack^2
\right. \nonumber \\ & & {\hskip -15.0truemm} \left. \phantom{12345}   
   \left\lbrack 10  e_i^2 \psi \psi_2+e_i^2 {\psi'}^2+16 e_r e_i \psi' \psi^2+4 (5 e_i^2+e_r^2) \psi^4\right\rbrack
\right\rbrace^2
+ \hbox{subdominant terms}=0.
\label{ODE3psi}
\end{eqnarray}
\medskip

As a by-product of the elimination process,
the rational expression (\ref{eqCGL5Phiprime}) of $\psi$ in terms of $M,M',M''$
is quite short,
while the rational expression of $M$ in terms of $\psi$ 
also involves the third derivative $\psi'''$ and is quite lengthy.
This is why the phase of the complex amplitude $A$ 
is qualified as a ``slave'' variable \cite{vS2003},
because it allows one to easily compute $\varphi$ from $M$
but not \textit{vice versa}.
\medskip

The purpose of this work is to explain on the above example 
how to overcome the difficulties
created by the consideration of $\psi$ in two specific methods.
\medskip

The paper is organized as follows.
In section \ref{sectionPZ}, as a prerequisite study,
we investigate the detailed structure of movable singularities
of $M$ and $\psi$,
some features of which had been previously overlooked.

In section \ref{section_Residues},
we apply a first method successively to $M$ and $\psi$
in order to build necessary conditions for $M$ or $\psi$ to be elliptic.
This proves easy for $M$ and sets up additional questions for $\psi$,
whose solution is provided.

In section \ref{sectionSubeqMethod},
we indicate how to correctly apply a second method 
(the subequation method \cite{MC2003,CMBook,CM2009}) to $M$ and $\psi$,
in order to build a first order ODE sharing elliptic or degenerate elliptic solutions 
with the above third order ODEs (\ref{eqCGL5Order3}) and (\ref{ODE3psi}).

In section \ref{section_Subeq_Laurent4},
we simply present the solution of CGL5 in which the square modulus $M$ is elliptic,
whose obtention by the subequation method is
detailed elsewhere \cite{ConteNgCGL5}.

% =========================================================
\section{Movable singularities of CGL5} 
\label{sectionPZ}

In this section,
we enumerate all the movable poles of either $M$ or $\psi$,
excluding those which represent a singular solution\footnote{
A singular solution \cite{ChazyThese} of an ODE is any solution 
which cannot be obtained from the general solution.
Such a solution must cancel an odd multiplicity factor
of the discriminant of the ODE.
For instance the ODE (\ref{eqCGL5Order3}),
whose discriminant is $G M^2 (e_i M^2 + d_i M -g_r)^2$, 
admits as singular solutions all those of $G=0$,
which must be rejected
since they are not solutions of the system (\ref{eqCGL5ReducRealSystem}).} 
of either the third order ODE (\ref{eqCGL5Order3}) for $M$
or the third order ODE (\ref{ODE3psi}) for $\psi$.
\medskip

The structure of singularities of (\ref{eqCGL5})
has been studied in \cite{MCC1994}.
A first type of singularity $\chi=\xi-\xi_0 \to 0$ is obtained
by balancing the terms $A_{xx}$ and $\mod{A}^4 A$
in (\ref{eqCGL5}),
\begin{eqnarray}
& &
A \sim A_0 \chi^{(-1/2+i \alpha)},\ 
\overline{A} \sim A_0 \chi^{(-1/2-i \alpha)},\ 
(-1/2+i \alpha) (-3/2+i \alpha) p + A_0^4 r=0,
\end{eqnarray}
these algebraic branch points of $A$
define four values of $A_0^2$ and two of $\alpha$,
\begin{eqnarray}
& &
(e_i A_0^4)^2 - 4 e_r A_0^4 -3=0,\
\alpha=\frac{e_i}{2} A_0^4,
\label{eqCGL5LeadingOrderA0}
\\ & &
A_0^2 =\varepsilon_2 \sqrt{\frac{2 e_r + \varepsilon_1 \Delta}{e_i^2}},\
\alpha=                    \frac{2 e_r + \varepsilon_1 \Delta}{2 e_i},\
\Delta=\sqrt{4 e_r^2 + 3 e_i^2},\
\varepsilon_1^2=\varepsilon_2^2=1.
\label{eqCGL5LeadingOrderReal}
\end{eqnarray}
At these singularities, 
the square modulus $M=\mod{A}^2$ displays four simple poles
\begin{eqnarray}
& &
M \sim m_0 \chi^{-1},\ m_0=A_0^2,
\label{eqCGL5_poles_M}
\end{eqnarray}
and a direct investigation of the third order ODE (\ref{eqCGL5Order3})
shows that $M$ admits no other movable poles\footnote{
The Laurent series $M=\pm (3/(4 e_r))^{1/2} \chi^{-1} + \cdots$ 
must be discarded since it 
cancels $G$ 
and thus represents a singular solution.}.
\medskip

Let us now count the poles of $\psi$ 
by considering the real system (\ref{eqCGL5ReducRealSystem}).
A first set of poles of $\psi$ arises from the simple poles of $M$, 
\begin{eqnarray}
& &
M=m_0 \chi_1^{-1}
\left[
1+\left(\frac{\csi}{4}+\frac{2 d_r m_0-2 e_i d_i m_0^3}{4(1+e_i^2 m_0^4)}\right) \chi_1
+\mathcal{O}(\chi_1^2)
\right],
\label{eqCGL5Laurent_poles_M}
\\ & &
\psi=\frac{e_i m_0^2}{2} \chi_1^{-1}
+ \frac{e_i m_0^2}{8}\csi + m_0 \frac{4 d_i +5 e_i d_r m_0^2- e_i^2 d_i m_0^4}{4(1+e_i^2 m_0^4)}
+\mathcal{O}(\chi_1),
\label{eqCGL5Laurent_poles_psi}
\end{eqnarray}
in which both invariances (\ref{eqInvariance_q})--(\ref{eqInvariance_csi})
require changing $m_0$ to $-m_0$.
This first set defines four different simple poles of $\psi$ when $q$ is nonzero
and only two when it vanishes
(in which case $M^2$ obeys an algebraic equation admitting two double poles).
\medskip

A second set of poles of $\psi$, not considered in \cite{MCC1994},
arises from the movable simple zeroes of $M$,
and this set is best computed from the system (\ref{eqCGL5ReducRealSystem}).
This is
\begin{eqnarray}
& &
\frac{1}{M}=\frac{1}{\Arbzero}\chi_2^{-1}
\left[
1 + \Arbone \chi_2
+\left\lbrace \Arbone^2 + \csi \Arbone - \frac{j}{3} g_r + \frac{2}{3} g_i  
\right\rbrace \chi_2^2
+\mathcal{O}(\chi_2^3)
\right],
\label{eqCGL5Laurent_zeroes_M}
\\ & &
\psi= \frac{j}{2} \chi_2^{-1}
\left[
1 
+ \left(\csi + \Arbone\right) \chi_2
+\left\lbrace \Arbone^2 + 2 \csi \Arbone + \frac{2}{3} g_i - \frac{4 j}{3} g_r +\frac{5}{6}\csi^2
\right\rbrace \chi_2^2
\right.
\nonumber \\ & &
\left. \phantom{12345}
+\frac{1}{2}\left\lbrace 
 (g_r + j g_i)\csi
 + \frac{3 j \csi^3}{4} 
-\frac{(3 d_i -j d_r)\Arbzero}{4} 
+ \frac{(11 j \csi^2 + 4 g_r + 4 j g_i)\Arbone}{4} 
\right. \right.
\nonumber \\ & &
\left. \left. \phantom{123456789}
+ 3 j \csi \Arbone^2
+ j \Arbone^3
\right\rbrace \chi_2^3
+\mathcal{O}(\chi_2^4)\right],
\label{eqCGL5Laurent_poles_psi_at_zeroes_M}
\end{eqnarray}
in which $\Arbzero$ and $\Arbone$ are arbitrary constants,
and $j$ is any square root of $-1$.
Again, both invariances (\ref{eqInvariance_q})--(\ref{eqInvariance_csi})
require changing $\Arbzero$ to $-\Arbzero$,
with the additional requirement $\Arbone=0$ when $\csi=0$.
This second set defines 
$2 N$ simple zeroes of $M$,
and
either $2 N$ (when $q\not=0$) 
or     $  N$ (when $q=0$) simple poles of $\psi$, with $N$ an undetermined integer.

\begin{remark}
In the nonlinear Schr\"odinger equation
($p$ real, $q$ real, $r=\gamma=0$),
the first integral $M \psi=\hbox{constant}$ implies that the only poles of $\psi$
arise from the zeroes of $M$.
\end{remark} 
\medskip

We have searched for possible additional poles of $\psi$
by directly investigating the third order ODE (\ref{ODE3psi})
for $\psi(\xi)$.
One thus finds 
three kinds of families of movable simple poles,
namely
the two previous kinds
(\ref{eqCGL5Laurent_poles_psi}) and
(\ref{eqCGL5Laurent_poles_psi_at_zeroes_M})
plus the following third kind comprising two families
\begin{eqnarray}
& &
\psi=p_0 \chi^{-1}
\left[
1 + \frac{\csi}{4} +\mathcal{O}(\chi)\right],\
4 (e_r^2 + 5 e_i^2) p_0^2 -16 e_r e_i p_0 + 21 e_i^2=0. 
\label{eqCGL5Laurent_poles_ODE3_of_psi}
\end{eqnarray}
However, 
this series 
cancels a factor of the discriminant of (\ref{ODE3psi})
having an odd multiplicity,
therefore it is a singular solution of (\ref{ODE3psi})
and it must be discarded
because it is not a solution of the system (\ref{eqCGL5ReducRealSystem}).
\medskip

Therefore another distinction between $M$ and $\psi$
is the difference of complexity of their singularity structure:
exactly 4 Laurent series for $M$,
an unknown number of series for $\psi$.
\medskip

\begin{remark}
Out of the two special values $q=0$ and $\csi=0$ allowing an invariance in the
differential system, see (\ref{eqInvariance_q})--(\ref{eqInvariance_csi}),
only the value $q=0$ is involved in the structure of singularities.
The value $\csi=0$ will show up in next sections \ref{section_Residues}
and \ref{section_Subeq_Laurent4}.
\end{remark}

% =========================================================
\section{On the characterization of nondegenerate elliptic solutions}
\label{section_Residues}
\medskip

This is a classical result that, inside a fundamental domain,
the sum of the residues of any elliptic function at its poles
is equal to zero.
This allowed Hone \cite{Hone2005} to take advantage of the Laurent series
to generate the following necessary conditions 
for the solution $u$ of an ODE to be nondegenerate elliptic,
\begin{eqnarray}
& & {\hskip -8.0 truemm}
\forall j \in \mathcal{N}, \forall k \in \mathcal{N}:\
C_{jk}^{u} \equiv
\sum_{\hbox{Laurent series}} \hbox{residue}\left(\left(u^{(k)}\right)^j\right) =0,
\label{eqConditionsResidues}
\end{eqnarray}
in which the sum extends to any subset of the set of Laurent series, 
and thus to isolate those parameters for which the ODE might have a nondegenerate elliptic solution.
\medskip

If one assumes that $M$ is elliptic, 
it follows from (\ref{eqCGL5Phiprime}) that $\psi$ is also elliptic,
so one can \textit{a priori} use the Laurent series of either $M$ or $\psi$ 
to generate necessary conditions for $M$ and $\psi$ to be elliptic.
However, as already noticed in section \ref{sectionPZ},
the situation is much more complicated with $\psi$ than with $M$.
Indeed, 
some Laurent series of $\psi$ (those near $\chi_2$) depend on arbitrary constants,
therefore one must first compute the number of different series 
(\ref{eqCGL5Laurent_poles_psi_at_zeroes_M}),
then one must also solve the generated residues conditions
for these extra arbitrary constants.
Let us apply the criterium successively to $M$ and to $\psi$.

% =========================================================
\subsection{Criterium of residues applied to $M$}
\label{section_residues_M}

The variable $M$ presents two advantages over $\psi$:
it admits exactly four Laurent series (\ref{eqCGL5Laurent_poles_M}),
and no arbitrary coefficient enters these series.
The sum (\ref{eqConditionsResidues}) can include one, two, three or four Laurent series.
However, with one or three series, the condition (\ref{eqConditionsResidues}) 
applied to $\left(M^{(0)}\right)^1$ generates $m_0=0$, which is forbidden.
With two series, because of the invariance $m_0 \to -m_0$ of (\ref{eqCGL5LeadingOrderA0}),
the generated conditions are identical to those with four series,
therefore only one case remains to study, that with four series in the sum 
(\ref{eqConditionsResidues}).
After computing the first seven terms of each series (\ref{eqCGL5Laurent_poles_M}),
one generates ten conditions $C_{jk}^{M}=0$, $jk=$ 01, 02, 03, 04, 05, 06, 07, 12, 13, 22,
\begin{eqnarray}
& & 
C_{01}^{M} =0,\
C_{02}^{M} \equiv \csi e_r=0,\
C_{03}^{M} \equiv \csi\left(19 e_r d_i(ei^2 + 16 e_r^2) - e_i d_r (9 ei^2+6 e_r^2)\right)=0,\
\nonumber \\ & &
C_{04}^{M} \equiv \csi P(e_r,e_i,d_r,d_i,g_r,g_i,\csi^2)=0,\
C_{12}^{M} \equiv \csi P(e_r,e_i,d_r,d_i,g_r,g_i,\csi^2)=0,\ \cdots
\label{eqConditionsResiduesM}
\end{eqnarray}
in which the two $P$'s are polynomials containing respectively 18 and 16 terms.
In the case $\csi\not=0$, the generated constraints are
\begin{eqnarray}
& & 
\csi\not=0:\
C_{02}^{M} \equiv e_r=0,\
C_{03}^{M} \equiv d_r=0,\
C_{04}^{M} \equiv 16 g_i + 3 \csi^2=0,\
C_{12}^{M} \equiv d_i=0.
\label{eqConditionsResiduesMcsi_nonzero}
\label{eqsub54constraints}
\end{eqnarray}
We will see in section \ref{section_Subeq_Laurent4} that these necessary conditions are sufficient
because there does exist an elliptic solution when they are obeyed.
\medskip

In the case $\csi=0$,
the parity invariance (\ref{eqInvariance_csi}) selects very few
nonidentically zero expressions $C_{jk}^{M}=0$, the first few ones being:
$C_{13}^{M}=P_{2,2,5,5,15,15}(g_r,g_i,d_r,d_i,e_r,e_i)$\footnote{
The indices denote the degree of $P$ in its arguments.}
     (140 terms, requires    6 terms in the series), 
$C_{32}^{M}$ (not computed,  8 terms in the series), 
$C_{15}^{M}$ (not computed, 10 terms in the series),
etc,
thus requiring the computation of many terms in the series
to generate the necessary conditions.
\medskip

Fortunately, 
the unpleasant feature of having to deal with the second set of poles $\chi_2$,
whose number is unknown and which adds arbitrary coefficients to the system
of necessary conditions,
can be avoided
by associating to $M$ a ``subdominant'' contribution of the ``slave variable'' $\psi$
in the following way.
Any product $\left( \psi^{(k_1)}\right)^{j_1} \left(M^{(k_2)}\right)^{j_2}$
which is holomorphic near $\chi_2$
enjoys the same properties as $M$,
i.e.~:
to have only four Laurent series at $\chi_1$,
to introduce no extra arbitrary coefficients.
\medskip

The case $\csi=0$ can then be settled easily, 
the simplest necessary conditions being,
\begin{eqnarray}
& & \left\lbrace
\begin{array}{ll}
\displaystyle{
\psi M:\ 
3 e_i (2 e_r^2+3 e_i^2) d_r - e_r (8 e_r^2+11 e_i^2) d_i=0,
}\\ \displaystyle{
\phantom{12345678901}
\hbox{solved as } d_r=\frac{e_r (8 e_r^2+11 e_i^2)}{3 e_i (2 e_r^2+3 e_i^2)},\ 
}\\ \displaystyle{
\psi M^2:\ 
32 (2 e_r^2 + 3 e_i^2)^2 e_i (3 e_i g_i + 4 e_r g_r) - (16 e_r^2+19 e_i^2)(4 e_r^2+3 e_i^2) e_r d_i^2=0,
}\\ \displaystyle{
\phantom{12345678901}
\hbox{solved as } g_i=
  -\frac{4 e_r}{3 e_i} g_r 
  +\frac{(16 e_r^2+19 e_i^2)(4 e_r^2+3 e_i^2)}{96 e_i^2 (2 e_r^2 + 3 e_i^2)^2} e_r d_i^2,\
}\\ \displaystyle{ 
\psi M^5:\   d_i e_r P_{2,4,12,14}(g_r,d_i,e_r,e_i)=0,\
}\\ \displaystyle{
\psi^3 M^3:\ d_i     P_{2,4,14,16}(g_r,d_i,e_r,e_i)=0,\ 
}\\ \displaystyle{ 
\psi M^6:\       e_r P_{3,6,18,21}(g_r,d_i,e_r,e_i)=0. 
}
\end{array}
\right.
\end{eqnarray}
With only seven terms in each series, 
this defines three sets of necessary conditions
\begin{eqnarray}
& &
\csi=0,\ e_r=\frac{3 \varepsilon}{2} e_i,\ d_r=\frac{29 \varepsilon}{15} d_i,\ 
g_r=-\frac{d_i^2}{5 e_i},\ g_i=-\frac{7 \varepsilon d_i^2}{e_i},\ 
\varepsilon^2=1,
\label{eqCGL5csi0_residuecond1}
\\ & &
\csi=0,\ d_i=d_r=g_i=g_r=0,
\label{eqCGL5csi0_residuecond2}
\\ & &
\csi=0,\ d_i=d_r=g_i=e_r=0,
\label{eqCGL5csi0_residuecond3}
\end{eqnarray}
the last set being the particular case $\csi=0$ of the conditions
(\ref{eqConditionsResiduesMcsi_nonzero}).
The first two sets (\ref{eqCGL5csi0_residuecond1})--(\ref{eqCGL5csi0_residuecond2})
could well be refined by using more terms in each series.

% =========================================================
\subsection{Criterium of residues applied to $\psi$} 
\label{section_residues_psi}

The motivation of this study comes from
a previous application of the criterium to $\psi$, 
in which Vernov \cite{VernovCGL5} found the set of necessary conditions
\begin{eqnarray}
& &
\csi=e_r=d_r=d_i=g_i g_r=0,
\label{eqCGL5_residuecond_Vernov}
\end{eqnarray}
which only detects the subcase $\csi=0$ of the elliptic solution 
(\ref{eqsub54mod}). 
This result was achieved from the Laurent series of $\psi^j$, $j=1,2,3,4$, 
but only from the series (\ref{eqCGL5Laurent_poles_psi}) near $\chi_1$,
discarding the second set of poles (\ref{eqCGL5Laurent_poles_psi_at_zeroes_M}) 
near $\chi_2$.
Let us show in this subsection
that, if one takes account of both sets of poles of $\psi$,
the single consideration of the variables $\left( \psi^{(k)}\right)^{j}$
allows one to recover the correct results.
\medskip

Since an elliptic function possesses as many zeroes as poles,
$M$ must have exactly four simple zeroes (\ref{eqCGL5Laurent_zeroes_M}).
The corresponding set of poles of $\psi$, necessarily simple,
are then the following:

\begin{enumerate}
\item --
($q\not=0$)
4 poles $\chi_1$ plus 
4 poles $\chi_2$; 

\item --
($q=0$)
2 poles $\chi_1$ plus 
2 poles $\chi_2$. 

\end{enumerate}

\medskip

Let us denote the triplets $(M_0,M_1,j)$ 
in (\ref{eqCGL5Laurent_poles_psi_at_zeroes_M})
as either the four sets
$(M_{0,1\pm}, M_{1,1\pm},\pm i)$,
\hfill\break\noindent
$(M_{0,2\pm}, M_{1,2\pm},\pm i)$ (case $q\not=0$),
or the two sets
$(M_{0,\pm}, M_{1,\pm},\pm i)$(case $q=0$).
\medskip

Let us denote $C_{jk}^{\psi P}$ the sum of the four residues 
associated to the four poles $\chi_1$ (\ref{eqCGL5Laurent_poles_psi}),
and $C_{jk}^{\psi Z}$ the sum of two residues 
associated to two of the poles $\chi_2$ (\ref{eqCGL5Laurent_poles_psi_at_zeroes_M}) with opposite values of $j$.

The first set of sum of residues evaluates to \cite{VernovCGL5}
\begin{eqnarray}
& & \left\lbrace
\begin{array}{ll}
\displaystyle{
C_{01}^{\psi P} \equiv      \frac{4 e_r}{e_i},\
C_{02}^{\psi P} \equiv \csi \frac{8 e_r^2+3 e_i^2}{2 e_i^2},\ 
}\\ \displaystyle{
C_{03}^{\psi P} \equiv   
  6 \frac{e_r}{e_i} g_i+4 \frac{e_r^2}{e_i^2} g_r 
 +\frac{9 e_i^2+32 e_r^2}{8 e_i^3}e_r \csi^2 
 + \frac{520 e_i^2 e_r^4+256 e_r^6+303 e_i^4 e_r^2+9 e_i^6}{64 e_i^3 (e_r^2 + e_i^2)^2} d_i^2
}\\ \displaystyle{ \phantom{12345}
  - \frac{(132 e_i^4+229 e_i^2 e_r^2+112 e_r^4)}{16 e_i^2 (e_r^2 + e_i^2)^2} e_r d_i d_r 
  +3 \frac{e_i^2 (64 e_r^4+135 e_i^2 e_r^2+81 e_i^4 )}{64 e_i (e_r^2 + e_i^2)^2} d_r^2,\
}\\ \displaystyle{
C_{12}^{\psi P} \equiv \csi \Big[
 g_i -2 \frac{e_r}{e_i} g_r - \frac{3}{16} \csi^2
 - \frac{112 e_i^4+ 23 e_i^2 e_r^2-224 e_r^4}{96 e_i^2 (e_r^2 + e_i^2)^2} d_i^2
}\\ \displaystyle{ \phantom{12345}
 + \frac{132 e_i^4+229 e_i^2 e_r^2+112 e_r^4}{32       (e_r^2 + e_i^2)^2} e_r d_i d_r 
 + \frac{ 19 e_i^4-123 e_i^2 e_r^2-232 e_r^4}{32 e_i   (e_r^2 + e_i^2)^2} d_r^2  
\Big],\
}\\ \displaystyle{
\dots,
}
\end{array}
\right.
\label{eqSum4Residuespsipoles}
\end{eqnarray}
and the second set 
to 
(denoting 
$\as   =M_{1,+}+M_{1,-}$, $\ad=M_{1,+}-M_{1,-}$),
\begin{eqnarray}
& & \left\lbrace
\begin{array}{ll}
\displaystyle{
C_{01}^{\psi Z} \equiv 0,\
C_{02}^{\psi Z} \equiv -\csi- \frac{\as}{2},\
C_{03}^{\psi Z} \equiv 
- g_r - \frac{3 i \ad}{4} (\as + 2 \csi),\ 
}\\ \displaystyle{
C_{12}^{\psi Z} \equiv 
\frac{1}{4} \as^3+\frac{3}{2} \csi \as^2+\frac{41}{16} \csi^2 \as
      +\frac{9}{8} \csi^3+\frac{3}{4}(2 \csi+\as) \ad^2 
}\\ \displaystyle{\phantom{12345}
      -i g_r \ad + g_i (2 \csi+\as)
      +\frac{d_r}{4}(M_{0,+}+M_{0,-})+\frac{3 i d_i}{4}(M_{0,+}-M_{0,-}),\
}\\ \displaystyle{
C_{04}^{\psi Z} \equiv 
\frac{5}{16} \as^3+\frac{15}{8} \csi \as^2+\frac{57}{16} \csi^2 \as
      +\frac{17}{8} \csi^3+\frac{15}{16}(2 \csi+ \as) \ad^2 
}\\ \displaystyle{\phantom{12345}    
      -\frac{5}{4} i g_r \ad + \frac{3}{4} g_i (2 \csi+\as)
      +\frac{d_r}{16}(M_{0,+}+M_{0,-})+\frac{3 i d_i}{16}(M_{0,+}-M_{0,-}),\     
}\\ \displaystyle{
\dots  
%C_{05}^{\psi Z} \equiv \hbox{27 terms},\ 
%C_{13}^{\psi Z} \equiv \hbox{37 terms},\ 
%C_{22}^{\psi Z} \equiv \hbox{61 terms},\ 
%C_{06}^{\psi Z} \equiv \hbox{65 terms},
%C_{07}^{\psi Z} \equiv \hbox{ terms}.
}
\end{array}
\right.
\label{eqSum4Residuespsizeroes}
\end{eqnarray}
\medskip

The sets of conditions to be solved are then 
(the arguments of $C_{jk}^{\psi Z}(\cdots)$ describe the additional unknowns)
\begin{eqnarray}
& 
q \not=0,\  N=2:\ &              C_{jk}^{\psi P} + C_{jk}^{\psi Z}(M_{0,1\pm}, M_{1,1\pm},\pm i)
                                                 + C_{jk}^{\psi Z}(M_{0,2\pm}, M_{1,2\pm},\pm i) =0,
\label{eqset4P4Z}
\\ & 
q     =0,\  N=1:\ &  \frac{1}{2} C_{jk}^{\psi P} + C_{jk}^{\psi Z}(M_{0,\pm}, M_{1,\pm},\pm i)=0.
\label{eqset2P2Z}
\end{eqnarray}
\medskip

The first set of conditions (\ref{eqset4P4Z}) 
contains too many unknowns to be solved when only seven terms
in each series are considered,
we leave the computation to the interested reader.
\medskip

For the second set of conditions (\ref{eqset2P2Z}),
since $q$ is zero,
the Laurent series 
(\ref{eqCGL5Laurent_zeroes_M})--(\ref{eqCGL5Laurent_poles_psi_at_zeroes_M})
must possess the invariance (\ref{eqInvariance_q}),
\begin{eqnarray}
& &
q=0:\ (M,\psi,\chi,M_0,M_1,j) \to (-M,\psi,\chi,-M_0,M_1,j),
\end{eqnarray}
and, when $\csi$ is also zero, the additional invariance (\ref{eqInvariance_csi}),
\begin{eqnarray}
& &
q=0,\ \csi=0:\ M_1=0,\ (M,\psi,\chi,M_0,j) \to (M,-\psi,-\chi,-M_0,j).
\end{eqnarray}
This second set of conditions evaluates to
\begin{eqnarray}
& & \left\lbrace
\begin{array}{ll}
\displaystyle{
\psi^1 :\ e_r=0,
}\\ \displaystyle{
\psi^2:\ 2 \as + \csi=0,
}\\ \displaystyle{
\psi^3:\ 8 g_r + 9 \csi \ad=0, 
}\\ \displaystyle{
\psi^5:\ 128 e_i (\ms) -9 i \csi \ad (43 \csi^2 + 272 g_i)=0,
}\\ \displaystyle{ 
\psi^4:\ \csi (3 \csi^2 + 16 g_i)=0,
}\\ \displaystyle{
{\psi'}^3 M^3:\ \ad (128 i e_i (\md) +1296 g_i \csi^2 + 279 \csi^4)=0,
}\\ \displaystyle{ 
\cdots
}
\end{array}
\right.
\end{eqnarray}
and only admits the two solutions
\begin{eqnarray}
& &  {\hskip -8.0 truemm}
\csi\not=0:\ e_r=0,\ g_i=-\frac{3}{16} \csi^2,\
M_{0\pm}^2= \frac{ \csi^2 (16  g_r \pm 9 i \csi^2)}{64 e_i},\
M_{1\pm}=\frac{(-9 \csi^2 \pm 16 i g_r)}{36 \csi},
\label{eqsol2P2Z}
\\ & &  {\hskip -8.0 truemm}
\csi=0:\ e_r=0,\ g_r=0,\ M_{1\pm}=0,\ M_{0+}^2+M_{0-}^2=0,
\label{eqsol2P2Z0}
\end{eqnarray}
therefore the subcase $\csi\not=0$ does succeed to generate the desired constraints
(\ref{eqsub54constraints}).
\medskip

\begin{remark}
The denominator $\csi$ in (\ref{eqsol2P2Z}) corresponds to the factor $\csi$
of the ${M'}^4$ term in (\ref{eqsub54psi}) and,
when $\csi$ vanishes, the singularities $\chi_2$ do not exist any more
in (\ref{eqsub54psi}), see (\ref{eqsubVernov}).
\end{remark}

When one now requires both sets of necessary conditions to hold true
(the set (\ref{eqConditionsResiduesMcsi_nonzero}),
(\ref{eqCGL5csi0_residuecond1})--(\ref{eqCGL5csi0_residuecond3})
generated in subsection \ref{section_residues_M},
and the above set (\ref{eqsol2P2Z})--(\ref{eqsol2P2Z0})),
then only two possibilities remain for an elliptic solution $M$ to exist:
either (\ref{eqsub54constraints}),
which does define a nondegenerate elliptic solution,
or $e_r=d_r=d_i=g_r=g_i=0$,
which defines the four rational solutions $M=m_0/(\xi-\xi_0), \psi=(e_i m_0^2/2)/(\xi-\xi_0)$.

% =========================================================
\section{Method to determine all elliptic and degenerate elliptic solutions}
\label{sectionSubeqMethod}

Consider an $N$-th order autonomous algebraic ODE (\ref{eqODE})
\begin{eqnarray}
& &
E(u^{(N)},...,u',u)=0,\
'=\frac{\D}{\D x},\
\label{eqODE}
\end{eqnarray}
admitting at least one Laurent series
\begin{eqnarray}
& &
u=\chi^p \sum_{j=0}^{+\infty} u_j \chi^j,\ \chi=x-x_0.
\label{eqLaurent}
\end{eqnarray}

There exists an algorithm \cite{MC2003} to find in closed form
all its elliptic or degenerate elliptic solutions.

Its successive steps are \cite{CM2009,CMBook}:
\begin{enumerate}
\item
Find the analytic structure of movable singularities
(e.g., 4 families of simple poles, 2 of double poles).
For each subset of families (e.g.~2 families of simple poles)
deduce the elliptic orders $m,n$ (e.g.~$m=2,n=4$) of $u,u'$
and perform the next steps.

\item
Compute slightly more than $(m+1)^2$ terms in the Laurent series.

\item
Define the first order $m$-th degree
subequation $F(u,u')=0$ 
(it contains at most $(m+1)^2$ coefficients $a_{j,k}$),
\begin{eqnarray}
& &
F(u,u') \equiv
 \sum_{k=0}^{m} \sum_{j=0}^{2m-2k} a_{j,k} u^j {u'}^k=0,\ a_{0,m}\not=0.
\end{eqnarray}
According to classical results of Briot-Bouquet and Painlev\'e 
(see details in \cite{CM2009}),
any elliptic or degenerate elliptic solution of (\ref{eqODE})
\textit{must} obey such an ODE,
which is called ``subequation'' of (\ref{eqODE}) because it is required
(in next step) to admit (\ref{eqODE})
as a differential consequence.

\item
Require each Laurent series (\ref{eqLaurent}) to obey $F(u,u')=0$,
\begin{eqnarray}
& & {\hskip -10.0 truemm}
F \equiv \chi^{m(p-1)} \left(\sum_{j=0}^{\jmax} F_j \chi^j
 + {\mathcal O}(\chi^{\jmax+1})
\right),\
\forall j\ : \ F_j=0.
\label{eqLinearSystemFj}
\end{eqnarray}
and solve this \textbf{linear overdetermined} system for $a_{j,k}$.

\item
Integrate each resulting ODE $F(u,u')=0$.
\end{enumerate}

The structure of singularities of $M$ and $\psi$ has been established 
in section \ref{sectionPZ},
and the result of concern to us is:
the movable poles of $M$ and $\psi$ are all simple,
and the number of distinct Laurent series at these simple poles
is equal to:
$4$ in the case of $M$,
$4+2 N$ in the case of $\psi$ or $M'/M$ if $q\not=0$,
and
$2+  N$ in the case of $\psi$ or $M'/M$ if $q=0$,
with $N$ an undetermined integer.
\medskip

Let us start with $M$.
In order to find all elliptic and degenerate elliptic solutions $M$
by the subequation method, 
at step 1 the various subsets of families to be considered are 
made of one, two, three or four series (\ref{eqCGL5Laurent_poles_M}),
defining subequations $F=0$ whose degrees in $(M',M)$ are respectively
$(1,2)$, $(2,4)$, $(3,6)$, $(4,8)$. 
The computation presents no other difficulties than technical ones
and its detailed results can be found in \cite{ConteNgCGL5}.
The main new result is a nondegenerate elliptic solution presented 
in section \ref{section_Subeq_Laurent4}.
\medskip

Applying the subequation method with $\psi$ presents several difficulties.
\begin{enumerate}
\item

The main one is not to forget the movable singularities of type $\chi_2$
(movable simple zeroes (\ref{eqCGL5Laurent_zeroes_M}) for $M$, 
 movable simple poles  (\ref{eqCGL5Laurent_poles_psi_at_zeroes_M}) for $\psi$).
This is the reason why a previous investigation \cite{VernovCGL5}
could only find a particular case of the elliptic solution
presented in section \ref{section_Subeq_Laurent4}.

\item

The number of Laurent series 
(\ref{eqCGL5Laurent_poles_psi_at_zeroes_M}) for $\psi$ is undetermined,
thus failing to set an upper bound to the degree of the subequation $F$.
This point has already been settled in section \ref{section_residues_psi},
where we have established that the number of distinct Laurent series $\psi$
is either eight ($q\not=0$, four series near $\chi_1$ plus four series near $\chi_2$)
or four ($q=0$, two series near $\chi_1$ plus two series near $\chi_2$).

\item

The arbitrary coefficients
in the series (\ref{eqCGL5Laurent_poles_psi_at_zeroes_M})
must be determined by the subequation method and therefore
require the computation of many more terms in each Laurent series.

\item

Another difficulty, undetectable \textit{a priori},
is the infinite value of $M_{1\pm}$ at $\csi=0$ in (\ref{eqsol2P2Z}).
Setting $\csi=0$ (for some reason) while looking for subequations
being obeyed by four series will fail.
In such a case ($q=0$ and $\csi=0$),
$\psi$ admits two distinct series near $\chi_1$ and no series near $\chi_2$,
and the correct assumption for $F$ is 
\begin{eqnarray}
& & {\hskip -10.0truemm}
F\equiv {\psi'}^2 + (a_{01} + a_{11}\psi + a_{21} \psi^2) \psi' 
+ a_{00} + a_{10}\psi + a_{20} \psi^2 + a_{30}\psi^3 + a_{40} \psi^4=0,
\end{eqnarray}
which is further reduced by a classical theorem
(no first degree term \cite[\S 181]{BriotBouquet} in $\psi'$ since $\psi$ is assumed elliptic)
and by the invariance (\ref{eqInvariance_csi})
to the canonical Briot-Bouquet type
\begin{eqnarray}
& & 
F\equiv {\psi'}^2 
+ a_{00}  + a_{20} \psi^2 + a_{40} \psi^4=0,
\end{eqnarray}
for which a solution was indeed found \cite{VernovCGL5}, see (\ref{eqsubVernov}).

\end{enumerate}

For all those reasons,
the choice of $M$ is by far the best one for amplitude equations
such as CGL3, CGL5 or others.

% =========================================================
\section{The elliptic solution of CGL5}
\label{section_Subeq_Laurent4}
\medskip

By ``elliptic solution of CGL5'', we mean a solution of CGL5 
in which the square modulus $M$ of the traveling wave reduction is elliptic.

The subequation method yields a unique genus-one subequation for $M$ 
\cite{ConteNgCGL5},
and it requires exactly the constraints (\ref{eqConditionsResiduesMcsi_nonzero}).
Introducing the shorthand notation,
\begin{eqnarray}
& & 
\ex=\frac{\csi^2}{48},\
\ey=\frac{g_r}{36},
\end{eqnarray}
this fourth degree subequation is
\begin{eqnarray}
& & \left\lbrace
\begin{array}{ll}
\displaystyle{
{M'}^4 
-2 \csi M {M'}^3
 +\frac{72}{e_i} \ex {M'}^2 (e_i M^2 - 12 \ey)
 +\frac{2^4 3^8 \ex^4}{e_i^2}
}\\ \displaystyle{\phantom{xxx}
 +\frac{648 \ex^2}{e_i^2} \left(288 \ey^2 +24 e_i \ey M^2 -e_i^2 M^4\right)
 - \frac{1}{3^4 e_i} M^2 \left(e_i M^2 -48 \ey\right)^3
=0.
}
\end{array}
\right.
\label{eqsub54Mexey}
\label{eqsub54mod}
\end{eqnarray}
\medskip

Because of the previously mentioned numerous difficulties with $\psi$,
the corresponding subequation for $\psi$ is not determined
by the subequation method,
but by elimination with the correspondence (\ref{eqCGL5Phiprime}),
resulting in
\begin{eqnarray}
& & \left\lbrace
\begin{array}{ll}
\displaystyle{  
  \csi {\psi'}^4 
- 4 \csi {\psi'}^3 \left(\csi \psi + 24 \ey\right)
}\\ \displaystyle{\phantom{xx}
+ 8 {\psi'}^2 \left(-\csi (27 \ex^2-324 \ey^2)
              +1440 \ex \ey \psi
              +27 \csi \ex \psi^2
              +16 \ey \psi^3
              +\frac{1}{3} \csi \psi^4\right)             
}\\ \displaystyle{\phantom{xx}   
 +16 \Big( 
              -\frac{1}{3} \csi \psi^8
              -\frac{32}{3} \ey \psi^7
              -26 \csi \ex      \psi^6
              -1632 \ex \ey \psi^5                 
             - \left(477 \ex^2+552 \ey^2 \right)\psi^4
     \Big.
}\\ \displaystyle{\phantom{xx} 
             - 288 \left(165 \ex^2+4 \ey^2\right)\psi^3                  
             +\csi \left(2106 \ex^2-31320 \ey^2\right)\psi^2
}\\ \displaystyle{\phantom{xx} 
\Big. 
             +2^7 3^6 \left(\ex^2-4 \ey^2\right) \ex \ey \psi    
             +243    \left(-9 \ex^4 +56 \ex^2 \ey^2 -144 \ey^4\right)  
\Big)=0.
}
\end{array}
\right.
\label{eqsub54psi}
\end{eqnarray}
\medskip

The singularity $\csi=0$ already uncovered in (\ref{eqsol2P2Z}) is displayed 
as the factor $\csi$ in front of the ${\psi'}^4$ term in (\ref{eqsub54psi}),
with the consequence that 
the degree of subequation (\ref{eqsub54psi}) drops from four to two
when $\csi=0$.
One then recovers the result of Vernov \cite{VernovCGL5},
\begin{eqnarray}
& & q=0,\ e_r=0,\ g_i=0,\ \csi=0,
\left\lbrace
\begin{array}{ll}
\displaystyle{
e_i (3 M')^4 - M^2 \left(3 e_i M^2 -4 g_r\right)^3=0,
}\\ \displaystyle{
9 {\psi'}^2 -12  \psi^4 - g_r^2=0,
}
\end{array}
\right.
\label{eqsubVernov}
\end{eqnarray}
in which both subequations for $M$ and $\psi$
belong to the list of five canonical equations of Briot and Bouquet.
\medskip

Full details on the integration of (\ref{eqsub54Mexey}) and (\ref{eqsub54psi})
can be found in \cite{ConteNgCGL5_details_wp}.
The final result
for the complex amplitude $A$ is,
\begin{eqnarray} 
& & {\hskip -18.0 truemm}
\forall \csi:\
A =\hbox{constant}\ e^{\displaystyle{-i \omega t + i \frac{c \xi}{2 p}}} \
\Elemsimp(\xi,-\xi_+^{\dlogA},0)^{(-1+i \sqrt{3})/2} 
\Elemsimp(\xi,-\xi_-^{\dlogA},0)^{(-1-i \sqrt{3})/2},
\label{eqAsigma}
\end{eqnarray}
in which $\Elemsimp(\xi,q,k)$ is the \textit{\'el\'ement simple} defined by Hermite
\cite[vol.~II, p.~506]{HalphenTraite}
for integrating the Lam\'e equation,
\begin{eqnarray}
& &
\Elemsimp(\xi,q,k)=\frac{\sigma(\xi+q)}{\sigma(\xi) \sigma(q)} e^{(k-\zeta(q)) \xi},
\label{eqdefElemsimp}
\end{eqnarray}
and the fixed constants $\xi_\pm^{\dlogA}$ in (\ref{eqAsigma}) are defined by
\begin{eqnarray}
& & {\hskip -18.0 truemm}
\left\lbrace
\begin{array}{ll}
\displaystyle{
\wp(\xi_\pm^{\dlogA},G_2,G_3)=
- 2 \ex \pm i \sqrt{3} (3 \ex + 4 i \ey),\ 
\wp'(\xi_\pm^{\dlogA},G_2,G_3)=
\frac{3 \mp i \sqrt{3}}{2}(3 \ex + 4 i \ey),
}\\ \displaystyle{
G_2=12 (13 \ex^2+16 \ey^2),\
G_3= 8 (35 \ex^2+48 \ey^2) \ex,
}\\ \displaystyle{
{\wp'}^2=4 (\wp+2 \ex) (\wp^2-2 \ex \wp-35 \ex^2-48 \ey^2).
}
\end{array}
\right.
\label{eqpolesdlogA}
\end{eqnarray}
\medskip

Numerical simulations with periodic boundary conditions 
\cite[Fig.~4]{PSAK}
do display solutions $M$ having a real period
(similar features are observed in CGL3 \cite[Fig.~7]{Chate1994}),
these could well correspond to the present elliptic solution.

% ==========================================================
\section{Conclusion}
\label{conclusion}

The traps described in this article should be kept in mind
when looking for all the elliptic or degenerate elliptic solutions
of other amplitude equations, such as the complex Swift-Hohenberg equation
\cite{SH1977}.

\section*{Acknowledgements}
RC warmly thanks the organizers for invitation,
and gladfully acknowledges the support of MPIPKS Dresden.
Part of this work was supported by
RGC under Grant No.~HKU 703807P.

% *************************************************************
% References


\begin{thebibliography}{99}

\bibitem{AK2002} I.S.~Aranson and L.~Kramer,
The world of the complex Ginzburg-Landau equation, 
Rev.~Math.~Phys.~{\bf 74} (2002) 99--143.
http://arXiv.org/abs/cond-mat/0106115

\bibitem{BriotBouquet} C.~Briot et J.-C.~Bouquet,        
{\it Th\'eorie des fonctions elliptiques},
1\`ere \'edition  \hfill\break\noindent
(Mal\-let-\-Bachelier, Paris, 1859);
2i\`eme \'edition (Gauthier-Villars, Paris, 1875).\hfill\break\noindent
\verb+http://gallica.bnf.fr/document?O=N099571+ 

\bibitem{Chate1994} H.~Chat\'e,                             
Spatiotemporal intermittency regimes of the one-dimensional complex
Ginzburg-Landau equation,
Nonlinearity {\bf 7} (1994) 185--204.

\bibitem{ChazyThese} J.~Chazy,                           
Sur les \'equations diff\'erentielles du troisi\`eme ordre et d'ordre
sup\'erieur dont l'int\'egrale g\'en\'erale a ses points critiques fixes,
Acta Math.~{\bf 34} (1911) 317--385.

\bibitem{CMBook} R.~Conte and M.~Musette,
{\it The Painlev\'e handbook} (Springer, Berlin, 2008).
Russian translation
{\it Metod Penleve y ego prilozhenia}
(Regular and chaotic dynamics, Moscow, 2011).

\bibitem{CM2009} R.~Conte and M.~Musette,
Elliptic general analytic solutions,
Studies in Applied Mathematics {\bf 123} (2009) 63--81.
http://arxiv.org/abs/0903.2009 

\bibitem{ConteNgODE3} R.~Conte and T.-W.~Ng, 
Meromorphic solutions of a third order nonlinear differential equation,
J.~Math.~Phys.~{\bf 51} (2010) 033518 (9 pp).

\bibitem{ConteNgCGL5} R.~Conte and T.W.~Ng,
to be submitted (2012).

\bibitem{ConteNgCGL5_details_wp} R.~Conte and T.W.~Ng,
Meromorphic traveling wave solutions of 
the complex cubic-quintic Ginzburg-Landau equation,
submitted (2011). 

\bibitem{HalphenTraite} G.-H.~Halphen,
{\it Trait\'e des fonctions elliptiques et de leurs applications}
(Gauthier-Villars, Paris, 1886, 1888, 1891). 
 \verb+http://gallica.bnf.fr/document?O=N007348+ 

\bibitem{Hone2005} A.N.W.~Hone,                                        
Non-existence of elliptic travelling wave solutions of the complex
Ginzburg-Landau equation,
Physica D {\bf 205} (2005) 292--306.

\bibitem{Klyachkin} A.V.~Klyachkin,
Modulational instability and autowaves in the active media described by the
nonlinear equations of Ginzburg-Landau type,
preprint 1339, Joffe, Leningrad (1989).

\bibitem{MCC1994} P.~Marcq, H.~Chat\'e and R.~Conte,
Exact solutions of the one-dimensional quintic complex Ginzburg-Landau
equation,
Physica D {\bf 73} (1994) 305--317. http://arXiv.org/abs/patt-sol/9310004

\bibitem{MC2003} M.~Musette and R.~Conte,
Analytic solitary waves of nonintegrable equations,
Physica D {\bf 181} (2003) 70--79.
http://arXiv.org/abs/nlin.PS/0302051

\bibitem{PSAK} S.~Popp, O.~Stiller, I.~Aranson, and L.~Kramer, 
Hole solutions in the 1d complex Ginzburg-Landau equation,
Physica D {\bf 84} (1995) 398--423.

\bibitem{vS2003} W.~van Saarloos,
Front propagation into unstable states,
Physics reports {\bf 386} (2003) 29--222.

\bibitem {SH1977} J.~Swift and P.C.~Hohenberg,                        
Hydrodynamic fluctuations at the convective instability,
Phys.~Rev.~A {\bf 15} (1977) 319--328.

\bibitem{VernovCGL5} S.Yu.~Vernov,                                
Elliptic solutions of the quintic complex one-dimensional Ginzburg-Landau
equation,
J.~Phys.~A {\bf 40} (2007) 9833--9844. 

\end{thebibliography}
\end{document}